\documentclass[12pt]{article}
\begin{document}
\title{Effect of the spherical Earth on a simple
pendulum\thanks{Published: Eur. J. Phys. {\bf 24} (2003) 125-130}}
\author {Lior M.~Burko
\\ Department of Physics
\\ University of Utah, Salt Lake City, Utah 84112}
\date{29 August, 2002}
\maketitle

\begin{abstract} 
We consider the period of a simple pendulum in the gravitational field of
the spherical Earth. Effectively, gravity is enhanced compared with the
often used flat Earth approximation, such that the period of the
pendulum is shortened. We discuss the flat Earth approximation, and show
when the corrections due to the spherical Earth may be of interest.
\end{abstract}

\section{Introduction}

The experimental fact that all objects fall with the same acceleration is
known as the Weak Equivalence Principle \cite{wheeler}. The first
systematic tests of the universality of free fall were done by Galileo
\cite{galileo}, who measured the acceleration of freely falling
objects, or of objects rolling down an inclined plane. Such an approach
suffers from great inaccuracies, which are related to the short time
scales involved. Indeed, Galileo was unable to accurately determine $g$,
the gravitational acceleration at the surface of the Earth, due
to the lack of
an accurate clock. Galileo
concluded that a much better way to check the universality of free fall
was to use a pendulum. Newton, and following him others,
improved on Galileo's experiments, and were able to determine $g$ quite
accurately. (Notice that it is much harder to measure $g$ than it is to
test the Weak Equivalence Principle.) Because the period is cumulative, by
measuring the time over many cycles of the pendulum one can increase the
accuracy in the measurement of $g$ significantly. Is the value of the
gravitational acceleration as determined by free-fall experiments 
identical to the value of the gravitational acceleration as
determined by the period of a pendulum? In this paper we shall study this
question, and show that the spherical geometry of the Earth affects the
{\it effective} gravitational acceleration differently in the two types
of experiments: the effective gravitational acceleration is greater than
$g$ for pendulum experiments, and smaller than $g$ for free-fall
experiments. 

The approximation that both Galileo and Newton used (and which is also
used in virtually all physics textbooks from the high school level to the
advanced undergraduate level \cite{gold,marion}), is to make the
following two assumptions on the Earth's gravitational field. First, one
assumes that the gravitational acceleration is independent of the
altitude above the surface of the Earth. That is, one assumes that during
its motion, a freely falling object or the pendulum's bob experiences a
constant acceleration due to gravity, such that the variation of the
gravitational acceleration with altitude (``vertical inhomogeneities'' 
of the gravitational field) is
neglected. Second, for the case of a pendulum, one assumes that the
gravitational field lines are parallel. That is, one neglects the
variations in the direction of the gravitational field (``lateral
inhomogeneities'' of the gravitational field) which come about because the
gravitational field lines
direct towards the center of the Earth. Effectively, this latter
approximation amount to adopting a model for the Earth wherein the Earth
is flat. We shall dubb this latter approximation henceforth as the ``flat
Earth'' (FE) approximation. The former approximation, when taken in tandem
with the FE approximation, is equivalent to assuming that the flat Earth
extends indefinitely is all directions. As in usual pendulum experiments
both assumptions are regularly made, we shall refer by the FE
approximation to both assumptions taken together. 

It turns out that the values of the gravitational acceleration as
determined by free fall or by using a pendulum are not identical if one
uses the flat Earth approximation (even without neglecting the vertical
inhomogeneities). Specifically, the
finite size of the spherical Earth acts in opposite ways: it effectively
{\it decreases} the gravitational acceleration in free fall experiments,
and effectively {\it increases} the gravitational acceleration in pendulum
experiments.  
Under most circumstances, the deviations of the effective gravitational
acceleration from $g$ due to the spherical Earth are tiny: the relative
change in $g$ is of order of the ratio of the length of the arm of the
pendulum to the radius of the Earth. For short pendula, the systematic
error involved in neglecting this effect is smaller than the systematic
errors in the neglections of other effects, notably the finite amplitude
effect, the mass of the wire which suspends the bob, the finite size of
the bob, decay of finite amplitudes, or the buoyancy effect
\cite{nelson}. Nelson and Olsson \cite{nelson} listed many systematics,
and included many sources for error which affect the measurement of $g$
with a pendulum only minutely. Even with a wire of moderate length (Nelson
and Olsson used a wire whose length was 3m), the FE approximation is
responsible for a systematic error which is considerably larger than some
of the systematics
which Nelson and Olsson did discuss. With longer arm, the spherical Earth
effect becomes even more important. As we shall see, for pendula of
lengths comparable with the usual length of the Foucault pendulum, the
spherical Earth effect is comparable to the finite amplitude effect. And yet, 
the effect of the spherical Earth is usually neglected in most treatments. 

The spherical Earth effect was treated briefly by Gough \cite{gough}, who
assumed an infinitely small amplitude. It turns out, however, that the
spherical Earth effect couples with finite amplitudes in a rather
complicated manner. 

In this paper we shall consider the effect of a spherical Earth on
the period of a simple pendulum, and show that this effect acts to shorten
the period. That is, when one uses the FE model and fits experimental data
to the parameters of that model, one gets a value for the gravitational
acceleration which is slightly higher than the actual one. As the FE 
approximation is used so frequently at all level of instruction  (from the
high school level to the advanced undergraduate level), it is instructive
to understand the nature of the FE approximation. In addition, the
analysis in the paper is appropriate for classroom instruction or as a
homework assignment at the intermediate or advanced undergraduate levels.  

\section{Model}
\begin{figure}
\input epsf
\centerline{ \epsfxsize 7.0cm
\epsfbox{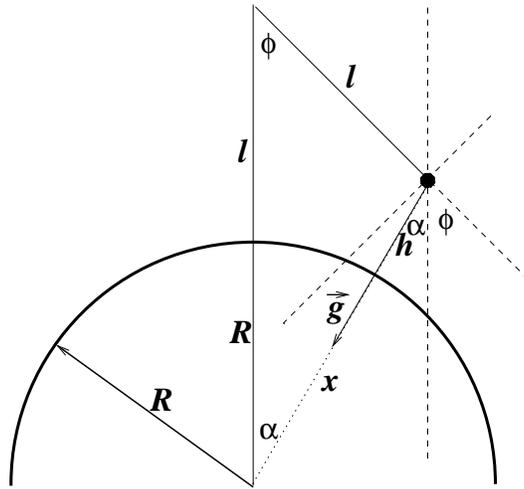}}
\caption{A pendulum with arm length $l$ is suspended above the Earth,
whose radius is $R$. The opening  angle of the pendulum's arm is 
$\phi$. The distance of the bob from the center of the Earth is $x$, and
from the surface of the Earth is $h$. The angle between the line which
connects the bob to the center of the Earth and the line which connects
the center of the Earth to the suspension point of the pendulum is
$\alpha$. The gravitational acceleration is ${\bf g}$.}
\label{fig1}  
\end{figure}

Consider a simple pendulum of arm length $l$, which moves under the
Earth's gravity, in the absence of any non-gravitational forces (such as
friction). We assume that the pendulum moves in a perfect vacuum, is
suspended from a perfectly rigid support by an inextensible massless
string, and that the bob is a point. We further assume that the Earth is
perfectly spherical. The radius of the Earth is $R$, and its mass is
$M$. At the surface of the Earth the gravitational acceleration then is
$g=GM/R^2$, where $G$ is Newton's constant. The opening angle of the
pendulum is $\phi$. (See Figure \ref{fig1}.) We assume that the arm of the
pendulum is extensionless and weightless, such that the entire mass of the
pendulum $m$ is in the bob. Denoting by $x$ the distance of the bob from
the center of the Earth, we find that
$x=\sqrt{(R+l)^2+l^2-2l(R+l)\cos\phi}$. 
The energy per unit mass ${\cal E}$ of the bob is 
\begin{equation}
{\cal E}=\frac{1}{2}l^2{\dot \phi}^2-\frac{gR^2}
{\sqrt{(R+l)^2+l^2-2l(R+l)\cos\phi}}\, ,
\end{equation}
where an overdot denotes differentiation with respect to time. 
Denoting $\beta\equiv l/R$, 
\begin{equation}
{\cal E}=\frac{1}{2}l^2{\dot \phi}^2-\frac{gl/\beta}
{\sqrt{1+2\beta(1+\beta)(1-\cos\phi})}\, .
\end{equation}
Conservation of energy implies that ${\cal E}$ equals at all times its
value at the turning point of the motion, where $\phi$ equals its maximum
value $\phi_0$. At the turning point $\dot\phi=0$, such that 
\begin{eqnarray}
\frac{1}{2}l^2{\dot \phi}^2&-&\frac{gl/\beta}
{\sqrt{1+2\beta(1+\beta)(1-\cos\phi})}\nonumber \\
=&-&\frac{gl/\beta}
{\sqrt{1+2\beta(1+\beta)(1-\cos\phi_0})}\, .
\end{eqnarray}

Define next $\psi\equiv\phi /2$, and find that 
\begin{equation}
2l^2{\dot \psi}^2-\frac{gl/\beta}
{\sqrt{1+\kappa\sin^2\psi}}
=-\frac{gl/\beta}
{\sqrt{1+\kappa\sin^2\psi_0}}\, ,
\end{equation}
where $\kappa=4\beta(1+\beta)$. Notice that for small values of $\beta$,
$\kappa$ is at order $\beta$.

\section{Determination of the period}

The period of the motion $T$ is given by
\begin{eqnarray}
T&=&4\int_0^{\psi_0}\frac{\,d\psi}{\dot\psi}\nonumber \\
&=&4\sqrt{\frac{2l\beta}{g}}\int_0^{\psi_0}\frac{\,d\psi}{
\left(\frac{1}{\sqrt{1+\kappa\sin^2\psi}}-
\frac{1}{\sqrt{1+\kappa\sin^2\psi_0}}\right)^{\frac{1}{2}}}\, .
\end{eqnarray}

This integral can be evaluated numerically. However, we can gain much
insight into it by approximating the integrand. As $\kappa\sin^2\psi_0\ll
1$, we can expand the expression inside the brackets as 
\begin{equation}
\frac{1}{\sqrt{1+\kappa\sin^2\psi}}-\frac{1}{\sqrt{1+\kappa\sin^2\psi_0}}
\approx \frac{\kappa}{2}(\sin^2\psi_0-\sin^2\psi)\, ,
\end{equation}
neglecting terms which are at order $\kappa^2\sin^4\psi_0$. 
Then, to that order, we find that 
\begin{equation}
T= 4\sqrt{\frac{l}{g}}\frac{1}{\sqrt{1+\beta}}
\int_0^{\psi_0}\frac{\,d\psi}{\sqrt{\sin^2\psi_0-\sin^2\psi}}\, .
\label{per}
\end{equation}

The integral is the usual complete elliptic integral (see appendix
\ref{appa}) \cite{arfken,grad}, such that 
\begin{equation}
T=4\sqrt{\frac{l}{g(1+\beta)}}K\left(\sin^2\psi_0\right)
\, .
\label{e8}
\end{equation}
Recall that this result is accurate only to $O(\kappa\sin^2\psi_0)$. We
write in Eq.~(\ref{e8}) the result in terms of the Elliptic integral just
for easy notation and comparison with the usual result of the FE
model. 
In the limit of $\beta\to 0$ (radius of the Earth goes to infinity, ``flat
Earth model"), we recover the regular result for the period of a simple
pendulum, i.e., $T_{\rm FE}=4\sqrt{l/g}K(\sin^2\psi_0)$. For
finite values of $\beta$, however, we find that the result can be
expressed through an effective gravitational acceleration
$g_*=g(1+\beta)$, which means that effectively gravity for a pendulum is
enhanced in the spherical Earth compared with the flat Earth model.
Consequently, we find that the period of a pendulum when
the spherical Earth is taken into consideration is shorter than in the 
flat Earth model. Specifically
\begin{equation}
T=T_{\rm FE}(1+\beta)^{-1/2}\, .
\label{period}
\end{equation}

We can also write the solution including higher-order corrections in
$\beta$. In this case one can no longer write the solution easily in
closed form. Instead, one can write it as a series expansion as 
\begin{equation}
T=4\sqrt{\frac{l}{g(1+\beta)}}\left[\frac{\pi}{2}+\frac{\pi}{8}
\psi_0^2(1+9\beta+9\beta^2)+O(\psi_0^4)\right]
\, .
\label{e9}
\end{equation}
Notice that the period in Eq.~(\ref{e8}) separated into a product of a
function of $\beta$ and a function of $\psi_0$. This happened just because
of the small $\kappa\sin^2\psi_0$ approximation we made. In general, the
period is not separable, as is clear from Eq.~(\ref{e9}). As the Elliptic
function $K(\sin^2\psi_0)=\pi/2+(\pi/8)\psi_0^2+O(\psi_0^4)$, in the limit
of $\beta\to 0$ Eq.~(\ref{e9}) indeed agrees with Eq.~(\ref{e8}).

Notice that the usual effect of finite amplitude is to lengthen the
period of the pendulum. The effect of the spherical Earth is in the
opposite direction: it shortens the period. This shortening of the period
comes about because effectively the gravitational acceleration is enhanced
compared with its value at the surface of the Earth. To understand
this heuristically, consider the forces which act on the bob. Because the
bob is at height $h$ above the surface of the Earth, the gravitational
acceleration there is given by $g_h=g(1+h/R)^{-2}$. Because the weight of
the bob directs towards the center of the Earth, its angle with the string
of the pendulum is not $\phi$, but rather $\phi+\alpha$ (see
Fig.~\ref{fig1} for the definitions of the geometrical
quantities.) Consequently, the projection of the bob's weight 
perpendicularly to the pendulum's arm, which is the force 
$f$ which determines the torque which drives the oscillations, is given by
\begin{equation}
f=mg\frac{\sin (\phi+\alpha )}{\left(1+\frac{h}{R}\right)^2}\, .
\end{equation}
Simple geometrical considerations show that 
$\sin\alpha =(l/x)\sin\phi$, and
$h=R(\sqrt{1+\kappa\sin^2\psi}-1)$. Expanding $f$ in small $l/R$, and
keeping only the leading terms in $\sin\psi$ and $l/R$, we find
that 
\begin{equation}
f=mg\left(1+\frac{l}{R}\right)\sin\phi=mg_*\sin\phi\, .
\end{equation}
This has the same form as the force which acts on the bob in the flat
Earth model, but with
a slightly enhanced gravitational acceleration. Gravity being effectively
stronger, it is clear why the period of the pendulum's oscillations
shortens. As the period of the pendulum is inversely proportional to the
square root of the gravitational acceleration, we immediately recover
Eq.~(\ref{period}).

In the flat Earth model we effectively take the limit as $l/R\to 0$. This
implies that when we compute the gravitational acceleration using a
pendulum, we in fact measure $g\left(1+\frac{l}{R}\right)$, not $g$, that
is, we measure a slightly larger acceleration. Of course, with reasonable
length for the pendulum, this effect is smaller than typical experimental
errors. Compare this with the gravitational acceleration which we measure
using a free-fall experiment (or an inclined surface). The gravitational
acceleration decreases with height, such that at all times it is slightly
smaller than $g$ at the Earth's surface. Specifically, the gravitational
acceleration is given by $g_h$. When we use the flat Earth (with infinite
extension) model, we in fact neglect this small variation in the
gravitational acceleration. That is, when measuring the gravitational
acceleration using a free fall experiment, we are in effect measuring
$g_h$, not $g$. Notice that $g_h<g$, whereas $g_*>g$. That is, when we use
the flat Earth model we measure different quantities for the gravitational
acceleration when we do a pendulum experiment or a free fall experiment. 

\section{Is the change of the pendulum's period anything but negligible?}

The fractional change in the pendulum's period being at order $l/R$
implies
that this effect is very small under most circumstances. As $R\approx
6\times 10^6$m, even if we take a long pendulum, of length $60$m, the
change in the period would be just one part in $10^5$. This effect, which
is typically ignored in textbooks, however, can be at the same order of 
magnitude as other effects, which many standard textbooks do
discuss \cite{gold,marion}. 
Specifically, one can choose reasonable parameters for which this
effect is at the same order as the usual lengthening of the period due to
finite amplitude effects. In fact, because these two effects are
competing, one can easily choose parameters for which the spherical Earth
effect compensates for the finite amplitude effect at leading order, such
that dependence on amplitude would be at order $\phi_0^4$ (rather than 
$\phi_0^2$). 

Comparing the two effects, we find that they are comparable if
$l/R\approx\phi_0^2/8$. Take, say, $\phi_0=6.24\times 10^{-3}$rad. The two
effects are comparable if $l/R\approx 4.87\times 10^{-6}$, or $l\approx
29.2$m. This length is the length of the Foucault pendulum in the Science
Museum of Virginia, and about the length of many Foucault's pendula  
which are used in many science museums for demonstration of the Earth's
rotation. The length of the arc of the pendulum's motion then is about
$0.364$m, which is large enough to be set conveniently. 

\section*{Acknowledgments}
I thank Richard Price for stimulating discussions. This research was
supported by the National Science foundation through Grant
No.~PHY-9734871.

\begin{appendix}

\section{Evaluation of the integral in Eq.~(\ref{per})} \label{appa}

In this appendix we  show how to bring the integral in Eq.~(\ref{per}) to
the familiar form of the complete elliptic integral of the first kind. The
strategy is to change the variables such that the interval of integration
would be from $0$ to $\pi/2$. We choose then a new variable $\varphi$,
defined by $\sin\psi=\sin\psi_0\sin\varphi$. The measure of the integral 
is
\begin{equation}
\,d\psi=\frac{\sin\psi_0\cos\varphi\,d\varphi}{\sqrt{1-
\sin^2\psi_0\sin^2\varphi}}\, .
\end{equation}
Substituting, we find that
\begin{equation}
\int_0^{\psi_0}\frac{\,d\psi}{\sqrt{\sin^2\psi_0-\sin^2\psi}}=
\int_0^{\pi /2}\frac{\,d\varphi}{\sqrt{1-\sin^2\psi_0\sin^2\varphi}}
\, ,
\end{equation}
which is nothing but the complete elliptic integral \cite{grad}  
$K(\sin^2\psi_0)$.

\end{appendix}

\end{document}